\RequirePackage[2020-02-02]{latexrelease}
\documentclass[%
 reprint,
superscriptaddress,
 amsmath,amssymb,
 aps,
 prd,
floatfix,
]{revtex4-2}
\usepackage{indentfirst}
\usepackage{graphicx}
\usepackage{dcolumn}
\usepackage{bm}
\usepackage{etoolbox}
\usepackage[export]{adjustbox}
\usepackage{mathtools}
\usepackage{float}
\DeclarePairedDelimiter{\ceil}{\lceil}{\rceil}
\usepackage{orcidlink}
\usepackage{aas_macros}
\usepackage{hyperref}
\hypersetup{colorlinks = true, citecolor = blue, linkcolor = blue}
\usepackage{bookmark}
\newtoggle{fullauthorlist}
\toggletrue{fullauthorlist}

\newtoggle{endauthorlist}

\begin{document}

\preprint{APS/123-QED}

\title{Search for gravitational waves from Scorpius X-1 with a hidden Markov model in O3 LIGO data}

\iftoggle{endauthorlist}{
  %
  %
  \let\mymaketitle\maketitle
  \let\myauthor\author
  \let\myaffiliation\affiliation
  \author{The LIGO Scientific Collaboration}
  \author{The Virgo Collaboration}
\noaffiliation
}{
  %
  %
  \iftoggle{fullauthorlist}{


\author{R.~Abbott}
\affiliation{LIGO Laboratory, California Institute of Technology, Pasadena, CA 91125, USA}
\author{H.~Abe}
\affiliation{Graduate School of Science, Tokyo Institute of Technology, Meguro-ku, Tokyo 152-8551, Japan  }
\author{F.~Acernese}
\affiliation{Dipartimento di Farmacia, Universit\`a di Salerno, I-84084 Fisciano, Salerno, Italy  }
\affiliation{INFN, Sezione di Napoli, Complesso Universitario di Monte S. Angelo, I-80126 Napoli, Italy  }
\author{K.~Ackley\,\orcidlink{0000-0002-8648-0767}}
\affiliation{OzGrav, School of Physics \& Astronomy, Monash University, Clayton 3800, Victoria, Australia}
\author{N.~Adhikari\,\orcidlink{0000-0002-4559-8427}}
\affiliation{University of Wisconsin-Milwaukee, Milwaukee, WI 53201, USA}
\author{R.~X.~Adhikari\,\orcidlink{0000-0002-5731-5076}}
\affiliation{LIGO Laboratory, California Institute of Technology, Pasadena, CA 91125, USA}
\author{V.~K.~Adkins}
\affiliation{Louisiana State University, Baton Rouge, LA 70803, USA}
\author{V.~B.~Adya}
\affiliation{OzGrav, Australian National University, Canberra, Australian Capital Territory 0200, Australia}
\author{C.~Affeldt}
\affiliation{Max Planck Institute for Gravitational Physics (Albert Einstein Institute), D-30167 Hannover, Germany}
\affiliation{Leibniz Universit\"at Hannover, D-30167 Hannover, Germany}
\author{D.~Agarwal}
\affiliation{Inter-University Centre for Astronomy and Astrophysics, Pune 411007, India}
% ...




  }{
    \author{The LIGO Scientific Collaboration}
    \affiliation{LSC}
    \author{The Virgo Collaboration}
    \affiliation{Virgo}
    \input{LSC-Virgo-KAGRA-Authors-Aug-2021}
  }
}

\date{\today}

\begin{abstract}
Results are presented for a semi-coherent search for continuous gravitational waves from the low-mass X-ray binary Scorpius X-1, using a hidden Markov model (HMM) to allow for spin wandering. This search improves on previous HMM-based searches of Laser Interferometer Gravitational-wave Observatory (LIGO) data by including the orbital period in the search template grid, and by analyzing data from the latest (third) observing run (O3).  In the frequency range searched, from 60 to 500 Hz, we find no evidence of gravitational radiation.  This is the most sensitive search for Scorpius X-1 using a HMM to date. For the most sensitive sub-band, starting at $256.06$Hz, we report an upper limit on gravitational wave strain (at $95 \%$ confidence) of $h_{0}^{95\%}=6.16\times10^{-26}$, assuming the orbital inclination angle takes its electromagnetically restricted value $\iota=44^{\circ}$. The upper limits on gravitational wave strain reported here are on average a factor of $\sim 3$ lower than in the O2 HMM search. This is the first Scorpius X-1 HMM search with upper limits that reach below the indirect torque-balance limit for certain sub-bands, assuming $\iota=44^{\circ}$. 

\end{abstract}

\maketitle

\section{\label{sec:INTRODUCTON} INTRODUCTION}

Rotating neutron stars are promising candidates for continuous-wave searches with terrestrial gravitational wave (GW) detectors such as the second-generation Advanced Laser Interferometer Gravitational-wave Observatory (LIGO) \cite{Riles2013,HarryLIGOScientificCollaboration2010,AcerneseAgathos2015,LIGOScientificCollaborationAasi2015,AnderssonFerrari2011}, Advanced Virgo \cite{LIGOScientificCollaborationAasi2015}, and the Kamioka Gravitational Wave Detector (KAGRA) \cite{KagraCollaborationAkutsu2019}. Continuous GWs from neutron stars are emitted by an oscillating quadrupole moment, which can be produced in various ways, including
 elastic strain \cite{UshomirskyCutler2000, Johnson-McDanielOwen2013},  magnetic gradients \cite{Cutler2002,MastranoMelatos2011,LaskyMelatos2013}, $r$-modes
 \cite{Heyl2002,ArrasFlanagan2003,BondarescuTeukolsky2009}, or nonaxisymmetric circulation of the superfluid interior \cite{PeraltaMelatos2006,vanEysdenMelatos2008,BennettvanEysden2010,MelatosDouglass2015}. These mechanisms emit GWs at specific multiples of the spin frequency $f_{\star}$ \cite{Riles2013}. Low-mass X-ray binaries (LMXBs) have been targeted by previous LIGO searches, \cite{AasiAbbott2015,WhelanSundaresan2015,AbbottAbbott2019a,MiddletonClearwater2020,ZhangPapa2021,TheLIGOScientificCollaborationtheVirgoCollaboration2021a} because they may emit GWs relatively strongly while existing in a state of rotational equilibrium, in which the accretion torque balances the GW torque \cite{PapaloizouPringle1978,Wagoner1984,Bildsten1998}. Under torque balance conditions, the characteristic GW strain $h_{0}$ is proportional to the square root of the X-ray flux, implying that the brightest LMXB, Scorpius X-1 (Sco X-1), is also a strong GW emitter \cite{Riles2013, Bildsten1998}.
 
Continuous-wave searches directed at Sco X-1 have been performed with data from the first (O1) and second (O2) observing runs of LIGO \cite{AasiAbbott2015,WhelanSundaresan2015,AbbottAbbott2017b,AbbottAbbott2017c,AbbottAbbott2017e,AbbottAbbott2019a,AbbottAbbott2019b2,ZhangPapa2021}. No signal has been detected to date, but astrophysically interesting upper limits have been obtained. For O1 a hidden Markov model (HMM) pipeline \cite{AbbottAbbott2017c} obtained an upper limit at $95\%$ confidence level of $h_{0}^{95\%}=8.3\times10^{-25}$ in the 100-200 Hz frequency range, while a cross-correlation (CrossCorr) pipeline \cite{DhurandharKrishnan2008,AbbottAbbott2017b} achieved $h_{0}^{95\%}=2.3\times10^{-25}$ in the same frequency range. For O2, the HMM pipeline \cite{AbbottAbbott2019a} obtained $h_{0}^{95\%}=3.47\times10^{-25}$, in the 100-200Hz frequency range, while CrossCorr \cite{ZhangPapa2021} improved on the O1 results by a factor of $\approx 1.8$. All of these upper limits on $h_{0}^{95\%}$ are marginalized over the neutron star spin inclination $\iota$, assuming an isotropic prior. If instead one assumes an electromagnetically informed prior,  $\iota=44^{\circ} \pm 6^{\circ}$ \cite{FomalontGeldzahler2001}, the O2 upper limits obtained by CrossCorr reduce to $\sim 10^{-26}$ \cite{ZhangPapa2021}.  Now, the third observation run (O3), which is longer and more sensitive than O1 and O2, offers an opportunity to repeat the O1 and O2 searches with improved sensitivity. The improved search is the subject of this paper. 

Searching for LMXBs presents two challenges. First, $f_{\star}$ (and hence the GW frequency) wanders stochastically in objects where it is measured electromagnetically, due to fluctuations in the hydromagnetic accretion torque \cite{deKoolAnzer1993,BaykalOegelman1993,BildstenChakrabarty1997,WattsKrishnan2008}. Second, in some LMXBs including Sco X-1, which do not exhibit X-ray pulsations, $f_{\star}$ is not measured electromagnetically \cite{WattsKrishnan2008,Riles2013}. Hence a GW search must cover a wide band (width $\sim 1$kHz) looking for an unknown, wandering, quasimonochromatic tone \cite{MukherjeeMessenger2018,TenorioKeitel2021}. HMM tracking is a tried and tested method for searches of the above sort, with a long history of practical use in telecommunications \cite{WhiteLBElliott} and remote sensing \cite{ParisSJauffretC}. HMM tracking has been used in numerous searches for GWs, e.g. for Sco X-1 in O1 \cite{AbbottAbbott2017c} and O2 \cite{AbbottAbbott2019a}, young supernova remnants in O3 \cite{TheLIGOScientificCollaborationtheVirgoCollaboration2021b}, accreting millisecond X-ray pulsars in O2 \cite{MiddletonClearwater2020} and O3 \cite{TheLIGOScientificCollaborationtheVirgoCollaboration2021a}, all sky searches \cite{BayleyMessenger2019}, and long-duration transients \cite{BanagiriSun2019}.
 
The $\mathcal{J}$-statistic \cite{SuvorovaClearwater2017}, a frequency domain matched filter, is used in tandem with the HMM described in references \cite{AbbottAbbott2017c,SuvorovaSun2016,SuvorovaClearwater2017,AbbottAbbott2019a}. The outline  of the rest of the paper is as follows. Section \ref{sec:SEARCH ALGORITHM} explains briefly the  HMM formulation used and the $\mathcal{J}$-statistic. In Section \ref{sec:O3 SEARCH} the search pipeline and parameter space are described. In Sections \ref{sec: O3 Analysis} and \ref{sec_V: Upper limits}, we present the search results and upper limits, respectively. We conclude in Section \ref{sec_VI:Conclusions}.

\section{\label{sec:SEARCH ALGORITHM} HMM ALGORITHM}

In Section \ref{subsecII:HMM FORMALISM} we review the HMM formalism used to track the wandering GW emission frequency from one time step to the next, according to a user-selected set of transition probabilities. For each time-step we calculate the likelihood of a signal being present given the data, via a maximum likelihood matched filter called the $\mathcal{J}$-statistic, which is reviewed in Section \ref{subesc_II:Jstat}.

\subsection{Hidden state structure and automaton}
\label{subsecII:HMM FORMALISM}

A HMM is a probabilistic finite state automaton characterized by a hidden state variable, $q(t)$, which takes the discrete values $\{q_{1},...\,,q_{N_{Q}}\}$, and an observable state variable, $o(t)$, which takes the discrete values $\{ o_{1},...\,,o_{N_{O}} \}$. The automaton jumps between states at discrete time epochs $\{ t_{1}, \, ... \, , t_{N_{T}} \}$. The probability of being in hidden state $q(t_{n+1})$ at time $t_{n+1}$ depends only upon the state at the previous time step $t_{n}$. This is known as a Markov process.

To complete the model two matrices are defined. First, the transition probability matrix, $A_{q(t_{n+1})q(t_{n})}$, which relates the probability of a state $q(t_{n})$ to jump to $q(t_{n+1})$, takes the form
\begin{align}
     A_{\scriptscriptstyle q(t_{n+1})q(t_{n})} = \frac{1}{3} &\Bigl[\left. \delta_{\scriptscriptstyle q(t_{n+1})q(t_{n})-\Delta q}+ \right. \nonumber \\
     &\left. \, \delta_{\scriptscriptstyle q(t_{n+1})q(t_{n})} + \delta_{\scriptscriptstyle q(t_{n+1})q(t_{n})+\Delta q} \right. \Bigr], \label{Eq_II_A:TransitionProb}
\end{align}

\noindent where $\delta_{ij}$ is the Kronecker delta. Eq.~\eqref{Eq_II_A:TransitionProb} defines the signal model as a piece-wise constant function that jumps by $-1, 0$, or $1$ frequency bin $\Delta q$ at each discrete transition time. See Section \ref{subsecIII:ScoX1params} for the details on the search frequency bin size. The other matrix is the emission probability matrix $L_{o(t_{n})q(t_{n})}$ that relates the likelihood of observing $o(t_{n})$ if the hidden state variable is $q(t_{n})$. $L_{o(t_{n})q(t_{n})}$ is constructed from the matched-filter $\mathcal{J}$-statistic, which we review in Section \ref{subesc_II:Jstat}. In our GW application, $f_{\star}(t)$ maps onto $q(t)$, noting that $f_{\star}(t)$ is hidden because it cannot be measured electromagnetically for Sco X-1. 

The data, specifically the Fourier transform of the time series output by the detector, map onto $o(t)$. The total observation time $T_{\text{obs}}$ is divided into $N_{T}$ segments of duration $T_{\text{drift}}$ rounded down to the nearest integer, i.e. $N_{T} = \lfloor T_{\text{obs}}/T_{\text{drift}} \rfloor$. $T_{\text{drift}}$ is chosen to prevent $f_{\star}(t)$ from wandering by more than one frequency bin per time step. The rate of spin wandering is unknown \textit{a priori} in Sco X-1, as $f_{\star}(t)$ cannot be measured electromagnetically. The observed X-ray flux variability can be used to estimate the stochastic variation in $f_{\star}(t)$ \cite{MessengerBulten2015,MukherjeeMessenger2018}, and from this we make an informed choice of $T_{\text{drift}}$, as in previous searches.

The probability that the hidden state follows some path $Q = \{q(t_{1}),...\,,q_{t_{N_{T}}}\}$ given some observed data $O = \{ o(t_{0}),\,...\,, o(t_{N_{T}}) \}$ is then given by the product of the transition and emission probabilities for each step, viz
\begin{align}
    P(Q\vert O) = \;\;\;&L_{o(t_{N_{T}})q(t_{N_{T}})}A_{q(t_{N_{T}})q(t_{N_{T}-1})}\,...\, \nonumber \\ 
    &\times L_{o(t_{1})q(t_{1})}A_{q(t_{1})q(t_{0})}\Pi_{q(t_{0})}, 
    \label{Eq_II_A:PQO}
\end{align}

\noindent where we define the prior probability of being in some initial state $q(t_{0})$ at time $t_{0}$ as $\Pi_{(t_{0})}$. In this paper we assign equal probability to all initial states, with $\Pi_{q(t_{0})} = 1/N_{Q}$.

We seek the path $Q^{*}$ that maximizes $P(Q\vert O)$ in Eq.~\eqref{Eq_II_A:PQO}. But it is computationally inefficient to consider all the $N_{Q}^{N_{T}+1}$ possible paths. A robust and computationally efficient way to find $Q^{*}$ is the Viterbi algorithm \cite{AViterbi1967}. This recursive algorithm exploits the principle of optimality to find $Q^{*}$ given $O$. A comprehensive description of the algorithm can be found in Appendix A of Ref.~\cite{AbbottAbbott2017c}.  

In this paper we use as detection statistic the log-likelihood of the most likely path $\mathcal{L} = \ln P(Q^{*} \vert O)$. 

\subsection{$\mathcal{J}$-statistic}
\label{subesc_II:Jstat}

The emission probability $L_{o(t_{n})q(t_{n})}$ relates the observed data, $o(t_{n})$, collected in the interval $t_{n} \leq t \leq t_{n}+T_{\text{drift}}$ to the hidden states $q(t_{n})$. In this paper, we express $L_{o(t_{n})q(t_{n})}$ in terms of the $\mathcal{J}$-statistic \cite{SuvorovaClearwater2017, AbbottAbbott2019a}. The $\mathcal{J}$-statistic is a maximum likelihood, frequency domain, matched-filter which tracks the orbital phase of the neutron star in its binary system. It is an extension of the traditional $\mathcal{F}$-statistic \cite{JaranowskiKrolak1998}, which is a matched filter for a biaxial rotor \cite{PrixKrishnan2009}. The Doppler shift due to the binary motion disperses the $\mathcal{F}$-statistic power into orbital sidebands of the GW carrier frequency. Although the $\mathcal{F}$-statistic can be used to produce matched filters to account for the Doppler shift due to the binary motion, the $\mathcal{J}$-statistic is used for computational efficiency. Section III in Ref.~\cite{SuvorovaClearwater2017} presents the detailed derivation of the $\mathcal{J}$-statistic.

In general, to account for the dispersed power, the $\mathcal{J}$-statistic is constructed from matched filters of the suggestive form
\begin{equation}
    G(f) = \mathcal{F}(f) \otimes B(f), 
    \label{Eq:Jstat}
\end{equation}

\noindent with 
\begin{equation}
    B(f) = \sum_{s = -m}^{m} J_{s}(2\pi f_{0} a_{0}) \exp(- \imath s \phi_{a})\delta\left(f-\frac{s}{P} \right),
    \label{eq:Jstatfilter}
\end{equation}

 \noindent where $\otimes$ denotes convolution, $J_{s}(x)$ is the Bessel function of the first kind of order $s$, with $m = {\rm ceil}\left( 2 \pi f_{0} a_{0} \right)$. Eq.~\eqref{eq:Jstatfilter} assumes the GW signal is produced by a biaxial rotor in a circular Keplerian orbit, and requires three binary orbital parameters: the projected semi-major axis $a_{0}$, the orbital phase at a reference time $\phi_{a}$, and $P$ is the binary orbital period.

\section{Search implementation} 
\label{sec:O3 SEARCH}

In this section we discuss the practical details of the search. Sections \ref{subsecIII:ScoX1params} and \ref{subsec_III:templates} define the parameter domain and grid respectively. Section \ref{subsecIII:Workflow} sets out the workflow. Section \ref{subsec_III:threshold} justifies the selection of the false alarm probability in terms of the number of search templates. Section \ref{subsec_III:Data} specifies the primary and secondary data products ingested by the search.

\subsection{Sco X-1 parameters}
\label{subsecIII:ScoX1params}

\begin{table*}
\caption{\label{tab:ScoX1Params} Search parameters and their range. The column headed ``EM data" records the availability of electromagnetic measurements in the references in the last column. All parameter ranges include the central value together with $\pm \,1\sigma$ uncertainties; values without uncertainties are treated as constants. In the text generally the central values are denoted with the subscript $0$, e.g. $P_{0}$. The time of ascension $T_{\rm asc}$ stands for the  value in \cite{WangSteeghs2018} propagated up to the start of O3, as described in Section \ref{subsecIII:ScoX1params}. As written here $T_{\rm asc}$ and $P$, plus their uncertainties, define a rectangular parameter domain. We only search over the grid points that have support from the propagated priors, defined by all the ellipses in the lower panel of Figure \ref{fig:propagation-Ellipses-example}; See Section \ref{subsec_III:templates} for details.}
\begin{ruledtabular}
\begin{tabular}{ccccc}
 Parameter & Symbol & Search range & EM data & Reference\\ \hline
 Right ascension& $\alpha$ & $16\,\text{h}\,19\,\text{m}\,55.0850\,\text{s}$ & Y &\cite{BradshawFomalont1999} \\
 Declination& $\delta$ & $-15^{\circ}38'24.9''$& Y & \cite{BradshawFomalont1999}\\
 Polarisation angle& $\psi$ & $234\,\pm\,3^{\circ}$ & Y & \cite{FomalontGeldzahler2001}\\
 Orbital inclination angle & $\iota$ & $44\,\pm6^{\circ}$ & Y & \cite{FomalontGeldzahler2001} \\
 Projected semi-major axis & $a_{0}$ & $2.35\pm0.3\,\text{s}$& Y & \cite{WangSteeghs2018}\\
 Orbital period& $P$ & $68\,023.86048\,\pm\,0.0432\,\text{s}$& Y & \cite{WangSteeghs2018}\\
 GPS time of ascension & $T_{\rm asc}$ & $1\,238\,149\,477.03488\pm\,200\,\text{s}$ & Y & \cite{WangSteeghs2018} \\
 Frequency& $f$ & $60\,-\, 500\,\text{Hz}$ & N & - \\
\end{tabular}
\end{ruledtabular}
\end{table*}

The $\mathcal{J}$-statistic depends on the direction of the source, described by the right ascension $\alpha$, and declination $\delta$. It also depends on the three binary orbital elements: $P$, $a_{0}$, and $\phi_{a}$. The time of passage through the ascending node, $T_{\text{asc}}$, is linked to $\phi_{a}$ via $\phi_{a} = 2 \pi T_{\text{asc}}/P \, (\text{mod} 2\pi)$. Henceforth we use $T_{\text{asc}}$ instead of $\phi_{a}$. Some of these parameters have been measured electromagnetically for Sco X-1. Their values and uncertainties are summarized in Table \ref{tab:ScoX1Params}. 

\begin{figure}
    \includegraphics[scale=0.28]{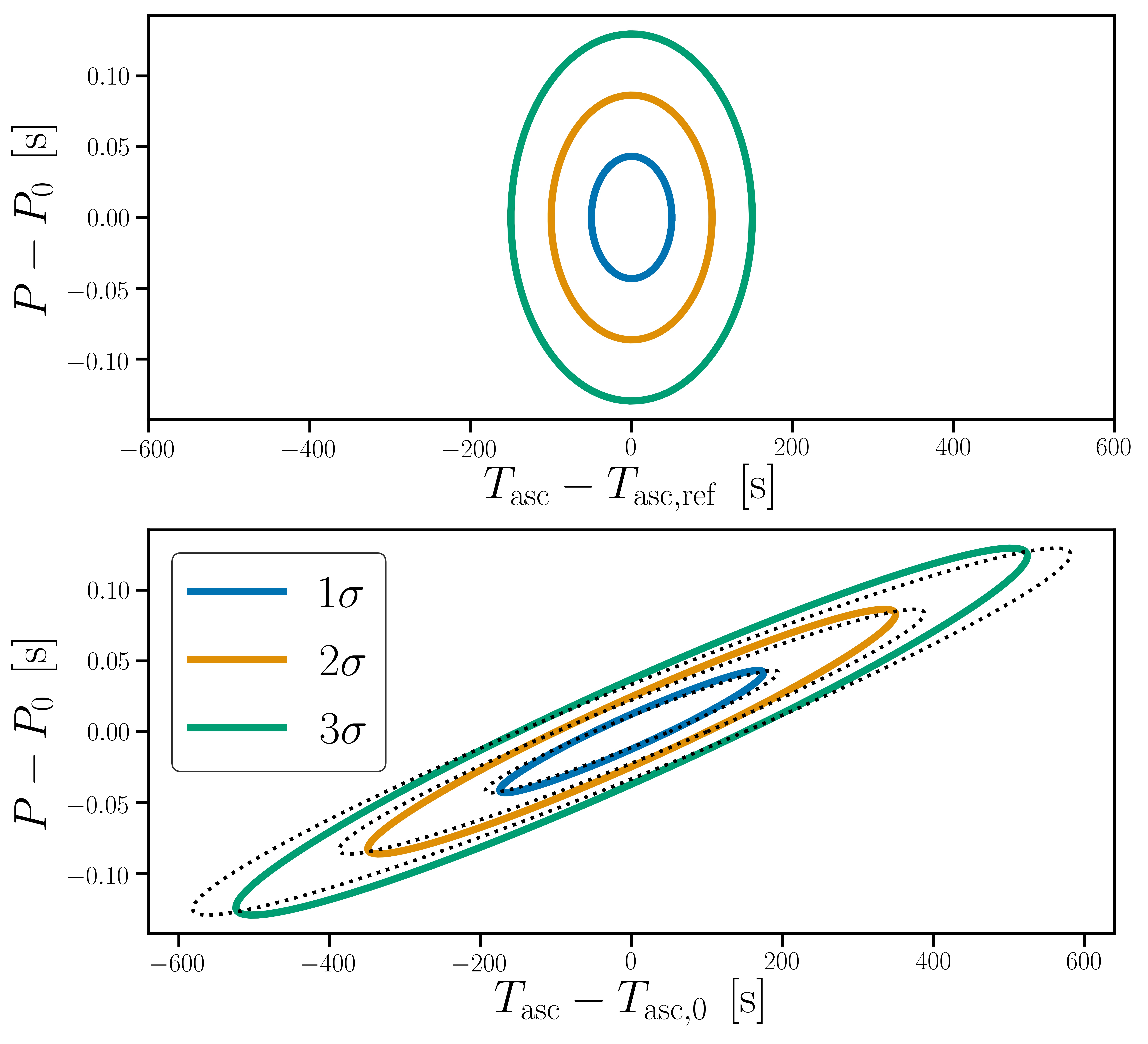}
    \caption{Uncertainty ellipses corresponding to $1\sigma$ [blue line], $2\sigma$ [yellow line], and $3\sigma$ [green line], for $T_{\rm asc}$ and $P$. The above panel show the curves at the reference time $T_{0}$, when $T_{\rm asc,ref}$ is uncorrelated with $P_{0}$ or its uncertainties. The second panel shows the effect of propagating $T_{\rm asc,ref}$ and $P_{0}$, with their respective uncertainties, to the start of O3 ($T_{\text{O}3,0}=1238166483$ GPS time) [solid lines] or the end of O3 ($T_{\rm O3 end}=1269361423$) [black dots]. $T_{\rm asc,0}$ represents the central value of $T_{\rm asc, ref}$ propagated to the start or end of O3; that is, $T_{\rm asc,0}$ is different for the solid and dotted lines. In both panels we have $P_{0}=68\,023.86048$s.} 
    \label{fig:propagation-Ellipses-example}
\end{figure}

The last electromagnetic measurements \cite{WangSteeghs2018} for the time of ascension $T_{\text{asc,ref}}=974\,416\,624$ GPS time are dated November 22 04:00:15 GMT 2010, here denoted $T_{0}$. We forward propagate $T_{\text{asc,ref}}$ to the start of O3, $T_{\text{O}3,0}=1238166483$ GPS time, viz.
\begin{equation}
    T_{\text{asc,0}} = T_{\text{asc,ref}}+N_{\text{orb}} P_{0},
    \label{Eq:Tasc_presentime}
\end{equation}

\noindent where $P_{0}=68\,023.86048$s is the central value of the orbital period in Table \ref{tab:ScoX1Params}, and $N_{\text{orb}}=\lceil\frac{T_{\text{O}3,0}-T_{0}}{P_{0}}\rceil$ is the number of full orbits between the reference time $T_{0}$ and $T_{\text{O}3,0}$. The original uncertainties for $T_{\text{asc,ref}}$ and $P$ are also propagated using Eq.~\eqref{Eq:Tasc_presentime}. This is illustrated in Figure \ref{fig:propagation-Ellipses-example}. The propagation maps the original uncertainty ellipses of $T_{\text{asc,ref}}$-$P$ (upper panel in Figure \ref{fig:propagation-Ellipses-example}) to the present uncertainty ellipses $T_{\text{asc}}$-$P$ (lower panel in Figure \ref{fig:propagation-Ellipses-example}). Propagating the original uncertainties via Eq.~\eqref{Eq:Tasc_presentime} creates correlations between the uncertainties of $T_{\text{asc}}$ and $P$.

The propagated priors on $T_{\rm asc}$ and $P$ change throughout the search duration as their correlation grows with time. The lower panel of Figure \ref{fig:propagation-Ellipses-example} shows the change from the start of O3 marked as solid color lines, to its end, $T_{\rm O3 end}=1269361423$ GPS time, shown as dotted lines. The search has been designed to cover the whole $3\sigma$ region of the propagated $T_{\rm asc}$-$P$ space, from start to end of O3.

For $a_{0}$, we cover the range $1.45 \leq a_{0}/(1\,\text{s}) \leq 3.25$, following the electromagnetic measurements  presented in \cite{WangSteeghs2018}. We can write this range equivalently as $\bar{a}_{0}\pm 3\sigma_{a_{0}}$, with $\bar{a}_{0}=2.35$s and $\sigma_{a_{0}}=0.3$s.

The coherence time is set to $T_{\text{drift}}=10 \; \text{d}$. The latter choice is justified astrophysically: it is the characteristic time-scale of the random walk in $f_{\star}$ inferred from accretion-driven fluctuations in the X-ray flux of Sco X-1 \cite{AasiAbbott2015,MessengerBulten2015,MukherjeeMessenger2018}. It also matches the value used in previous published searches for Sco X-1, enabling direct comparison with historical results \cite{AasiAbbott2015, AbbottAbbott2017b,AbbottAbbott2017c,AbbottAbbott2019a}. The resolution in frequency space $\Delta f_{\rm drift}$, i.e the size of the frequency bins, is set by the coherence time as $\Delta f_{\rm drift} = 1/(2 T_{\rm drift})=5.787037\times10^{-7} \rm Hz$. 

The range of frequencies to be searched, $60$ Hz to $500$ Hz, covers the region where LIGO is most sensitive. This frequency range is divided into sub-bands. This eases the manipulation of the data and allows one to approximate $f_{0}$ in Eq.~\eqref{eq:Jstatfilter} by the mid-point frequency $\bar{f}$ in each sub-band, accelerating the process of creating each matched filter via Eq.~\eqref{eq:Jstatfilter} \cite{SuvorovaClearwater2017}. The sub-bands are designed to contain a number of frequency bins $N_{f}$ that is a power of two, in order to accelerate the fast Fourier transform involved in calculating the convolution in Eq.~\eqref{eq:Jstatfilter}. For this search we use $N_{f} = 2^{20}$, setting the sub-band width to $\Delta f_{\text{sub}} = N_{f} \Delta f_{\text{drift}} = 0.6068148 \; \text{Hz}$. As such the total number of sub-bands to consider in our search is  $N_{\text{sub}} = \lceil (500-60) \text{Hz}/ \Delta f_{\text{sub}} \rceil= 725$. As the $\mathcal{J}$-statistic is less sensitive farther away from $\bar{f}$, we create sub-bands  with an overlap of $\Delta f_{\rm sub}/4$. This way the area with less sensitivity in a sub-band is covered with greater sensitivity in the neighbouring sub-band. All of the parameters discussed in this section are summarized in Table \ref{tab:ScoX1Params}.

\subsection{Number and placing of orbital templates}
\label{subsec_III:templates}

In this subsection we describe the procedure to calculate and place the orbital templates needed to cover the search parameter domain per sub-band. 

As explained in Section \ref{subsecIII:ScoX1params}, the forward propagation of the reference $T_{\rm asc, ref}$ increases its uncertainty as a function of $N_{\rm orb}$. To be conservative, we choose to propagate the uncertainties on $T_{\rm asc, ref}$ to the end of O3 with 
\begin{equation}
    \sigma_{T_{\rm asc}} = \sqrt{\sigma_{T_{\rm asc, ref}}^{2}+(N_{\rm orb, end}\sigma_{P})^{2}},
    \label{eq:uncertainty_Tasc}
\end{equation}

\noindent where one has $N_{\rm orb, end}=\lceil\frac{T_{\rm O3 end}-T_{0}}{P_{0}}\rceil$, and $\sigma_{T_{\rm asc, ref}}=50$s and $\sigma_{P}=0.0432$s are the uncertainties for $T_{\rm asc, ref}$ and $P$ respectively, the result $\sigma_{T_{\rm asc}}=200$s is included in Table \ref{tab:ScoX1Params}. Although the uncertainties are propagated to the end of O3, the central value for the time of ascension is only propagated to the start of O3, $T_{\rm asc_{0}}=1\,238\,149\,477.03488$ s, as shown in Section \ref{subsecIII:ScoX1params}.

We cover the parameter space by using a rectangular grid defined by the limits $(\bar{a}_{0}\pm 3\sigma_{a_{0}},T_{\rm asc}\pm 3\sigma_{T_{\rm asc_{0}}},P_{0}\pm3\sigma_{P})$. We set the spacing of the grid points  by selecting an acceptable maximum mismatch $\mu_{\rm max}$, following \cite{LeaciPrix2015}. This mismatch represents the fractional loss in signal-to-noise ratio between the search with the true parameters and the nearest grid point. For the search we use $\mu_{\rm max}=0.1$. The number of grid points needed to cover the $(a_{0},T_{\rm asc},P)$ space, for a given $\mu_{\rm max}$, are calculated using Eq.~(71) of Ref.~\cite{LeaciPrix2015}, namely
\begin{equation}
    N_{a_{0}} = \ceil[\Bigg]{ 3 \sqrt{2} \mu_{\rm max}^{-1/2} f \sigma_{a_{0}} },
    \label{eq:Numbera0}
\end{equation}

\begin{equation}
    N_{T_{\rm asc}} = \ceil[\Bigg]{ 6 \pi^{2} \sqrt{2} \mu_{\rm max}^{-1/2} f a_{0} P^{-1} \sigma_{T_{\rm asc}}},
    \label{eq:NumberTasc}
\end{equation}

\begin{equation}
    N_{\text{P}} = \ceil[\Bigg]{\pi^{2}\sqrt{6}\mu_{\rm max}^{-1/2}f a_{0} T_{\text{drift}} N_{T}P^{-2}\sigma_{P}}.
    \label{eq:NumberP}
\end{equation}

For O3 the number of contiguous semi-coherent segments is $N_{T}=36$. To be conservative when applying Eqs.(\ref{eq:Numbera0})-(\ref{eq:NumberP}), we use the highest frequency in each sub-band for $f$, the highest $a_{0}=\bar{a}_{0}+3\sigma_{a_{0}}$ and the lowest $P=P_{0}-3\sigma_{P}$. Table \ref{tab:guided_freqs_and_Ns} lists the number of grid points in selected sub-bands. For $T_{\rm asc}$ and $P$ we search only the points defined by Eqs.(\ref{eq:Numbera0})-(\ref{eq:NumberP}) that lie within the start (color lines) and end (dotted lines) ellipses in the bottom panel of Figure \ref{fig:propagation-Ellipses-example}.\\

\begin{table}
\begin{ruledtabular}
\begin{tabular}{cccc}
Sub-band (Hz) & $N_{a_{0}}$ & $N_{T_{\rm asc}}$ & $N_{P}$ \\ \hline
$60$ & $767$ & $149$ & $5$ \\
$160$ & $2031$ & $394$ & $12$ \\
$260$ & $3296$ & $640$ & $19$ \\
$360$ & $4560$ & $885$ & $27$ \\
$460$ & $6331$ & $1228$ & $37$ \\
\end{tabular}
\end{ruledtabular}
\caption{Selected sub-bands and corresponding $N_{a_{0}}, N_{T_{\rm asc}}, N_{\text{P}}$ values, using $\mu_{\rm max}=0.1$. The table serves to illustrate how the number of orbital templates varies across the full frequency band.}
\label{tab:guided_freqs_and_Ns} 
\end{table}

\subsection{Workflow}
\label{subsecIII:Workflow}

\begin{figure}[ht]
    \centering
    \includegraphics[width=6.9cm,keepaspectratio]{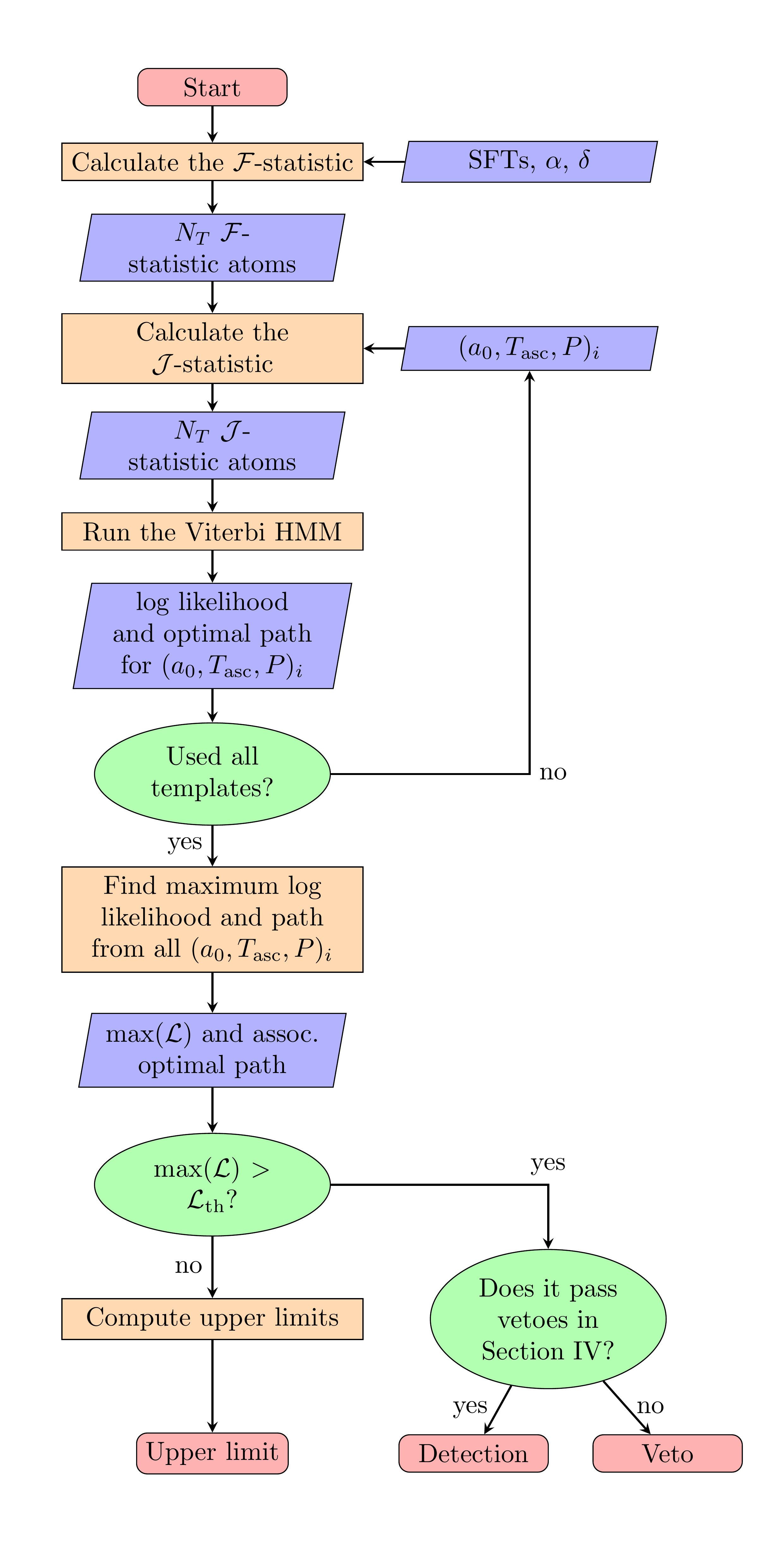}
    \caption{Flowchart of the search pipeline for a sub-band. The light red octagons are the start and end points, the orange rectangles are processes, the blue parallelograms are input or output data, and the green ovals stand for decision points. The full search repeats the steps in the flowchart for $725$ sub-bands from $60$Hz to $500$Hz.}
    \label{fig:flowchart}
\end{figure}

The workflow for the search is illustrated in Figure \ref{fig:flowchart}, as a flow chart. 

At the outset, the time series from the detector is converted into short  Fourier transforms (SFTs) lasting  $T_{\text{SFT}} = 1800$ s. The corresponding data, for each frequency sub-band, are divided into $N_{T}$ blocks of duration $T_{\rm drift}=10 ~\rm d$. All of the SFTs in a single block are used to calculate an $\mathcal{F}$-statistic \textit{atom} \cite{Fatoms} with the fixed parameters $\alpha$ and $\delta$ in Table \ref{tab:ScoX1Params}. The process is repeated for all the blocks, generating $N_{T}$ atoms. The $\mathcal{F}$-statistic atoms  do not depend on the orbital parameters so they are stored in a look-up table. The steps in this paragraph conclude in the blue parallelogram denoted ``$N_{T} \; \mathcal{F}$-statistic atoms" in Figure \ref{fig:flowchart}.

In every sub-band, each $\mathcal{F}$-statistic atom is fed into the $\mathcal{J}$-statistic in Eq.~\eqref{eq:Jstatfilter} for a triad of orbital parameters $(a_{0}, T_{\rm asc}, P)$. The Viterbi algorithm (see \cite{AViterbi1967} or Section II.B of Ref.~\cite{SuvorovaClearwater2017}) finds the optimal frequency path connecting the $\mathcal{J}$-statistic blocks and its associated log likelihood. In Figure \ref{fig:flowchart}, the latter steps extend from the orange rectangle ``Calculate the $\mathcal{J}$-statistic" up to the blue parallelogram marked ``log likelihood and optimal path for $(a_{0}, T_{\rm asc}, P)_{i}$".

For a sub-band centered on the frequency $\bar{f}$, the search scans over all binary templates $(a_{0}, T_{\rm asc}, P)_{i}$ with $1 \leq i \leq N_{a_{0}}(\bar{f})N_{T_{\rm asc}}(\bar{f})N_{P}(\bar{f})$, calculated using Eqs.(\ref{eq:Numbera0})-(\ref{eq:NumberP}). This step is illustrated as the green oval denoted ``Used all templates?" in Figure \ref{fig:flowchart}. For each $(a_{0}, T_{\rm asc}, P)_{i}$ an optimal path and its associated log likelihood are recorded. 

Following the loop over all binary templates, the optimal path with highest log likelihood, denoted $\rm max(\mathcal{L})$, is selected in the blue parallelogram marked ``$\rm max(\mathcal{L})$ and assoc. optimal path" in Figure \ref{fig:flowchart}. If $\rm max(\mathcal{L})$ is higher than the detection threshold  (see Section \ref{subsec_III:threshold}), then the sub-band is recorded as a candidate and passed through a hierarchy of vetoes (see Section \ref{subsec_IV:Vetoes}), via the ``yes" output of the upper green oval. 
 
Sub-bands with $\rm max(\mathcal{L})$ below the detection threshold are used to calculate GW strain upper limits via the ``no" output of the upper green oval. Vetoed sub-bands are not used to calculate GW upper-limits.

\subsection{False alarm probability and detection threshold}
\label{subsec_III:threshold}

A sub-band is registered as a candidate, when ${\rm max(\cal L)}$ exceeds a threshold, ${\cal L}_{\rm th}$, corresponding to a user-selected false alarm probability. As the distribution of $\rm max(\mathcal{L})$ in pure noise is unknown we rely on Monte-Carlo simulations to determine $\cal L_{\rm th}$ in each sub-band of the search. 
To estimate the distribution of $\rm max(\mathcal{L})$ in pure noise, we generate synthetic Gaussian data using the \texttt{lalapp\_Makefakedata\_v5} program in the LIGO Scientific Collaboration Algorithm Library (LALSuite) \cite{lalsuite}. The synthetic data are generated for a sub-band centered on the frequency $\bar{f}$, with $\alpha$ and $\delta$ copied from Table \ref{tab:ScoX1Params}. Then the search workflow described in Section \ref{subsecIII:Workflow} is applied. To avoid needless computation, we limit the grid to $N_{a_{0}}=322, N_{T_{\rm asc}}=35,$ and $N_{P}=5$ in every sub-band, independent of $\bar{f}$.

In general $\mathcal{L}_{\rm th}$ depends on the number of generated log likelihoods per sub-band, i.e. $N_{\rm tot}=N_{f}\,N_{a_{0}}\,N_{T_{\rm asc}}\,N_{P}$. We describe the false alarm probability, $\alpha_{N_{\rm tot}}$, of a sub-band with $N_{\rm tot}$ log likelihoods in terms of the probability of a false alarm in a single terminating frequency bin per orbital template, $\alpha$, as
\begin{equation}
    \alpha_{N_{\rm tot}} = 1 - ( 1 - \alpha )^{N_{\rm tot}}.
    \label{Eq:falsealarm_Ntot}
\end{equation}
 
Historically HMM Sco X-1 searches \cite{AbbottAbbott2017c,AbbottAbbott2019a} use $\alpha_{N_{\rm tot}}=0.01$, a choice we adopt here. A false-alarm probability of $1\%$ per sub-band applied to a total of $725$ sub-bands implies we should expect $\sim 7$ false alarms from the search. Searches with electromagnetically constrained $f_{\star}$ such as Ref.~\cite{MiddletonClearwater2020} and Ref.~\cite{TheLIGOScientificCollaborationtheVirgoCollaboration2021a} allow for  $\alpha_{N_{\rm tot}}=0.3$, given the reduced search space. Appendix A in Ref.~\cite{TheLIGOScientificCollaborationtheVirgoCollaboration2021a} presents the detailed procedure to set thresholds using Monte-Carlo simulations.  

\subsection{O3 data}
\label{subsec_III:Data}

The search uses all the O3 dataset, starting April 1, 2019, 15:00 UTC and finishing March 27, 2020, 17:00 UTC. The dataset is divided in two. The first part (O3a) spans 1 April 2019 to 1 Oct 2019 followed by a month long commissioning break. The second part (O3b) was intended to span 1 November 2019 to 30 April 2020 but was suspended in March 2020 due to the COVID-19 coronavirus pandemic. SFTs are generated from the ``C01 calibrated self-gated" dataset, specifically designed to remove loud glitches from the strain data, following the procedure in Ref.~\cite{CalibrationO3}.

Due to the month-long commissioning break between O3a and O3b, two out of $N_{T}$ segments have no SFTs. The two segments are dated October 8, 2019, 15:00 UTC and October 15, 2019, 15:00 UTC, respectively. We replace them by segments with uniform log-likelihood across all frequency bins, to allow the HMM to connect data from O3a with data from O3b while accommodating spin wandering. Every time atoms are created, such as in Sections \ref{subsec_III:threshold}, \ref{veto:single_IFO} and \ref{subsec_IV:Vetoes}, the relevant missing atoms are replaced with uniform log-likelihood atoms.

\section{ O3 Analysis}
\label{sec: O3 Analysis}

\subsection{Candidates}
\label{subsec:Candidates}

\begin{figure*}[ht]
    \centering
    \includegraphics[scale=0.410, left]{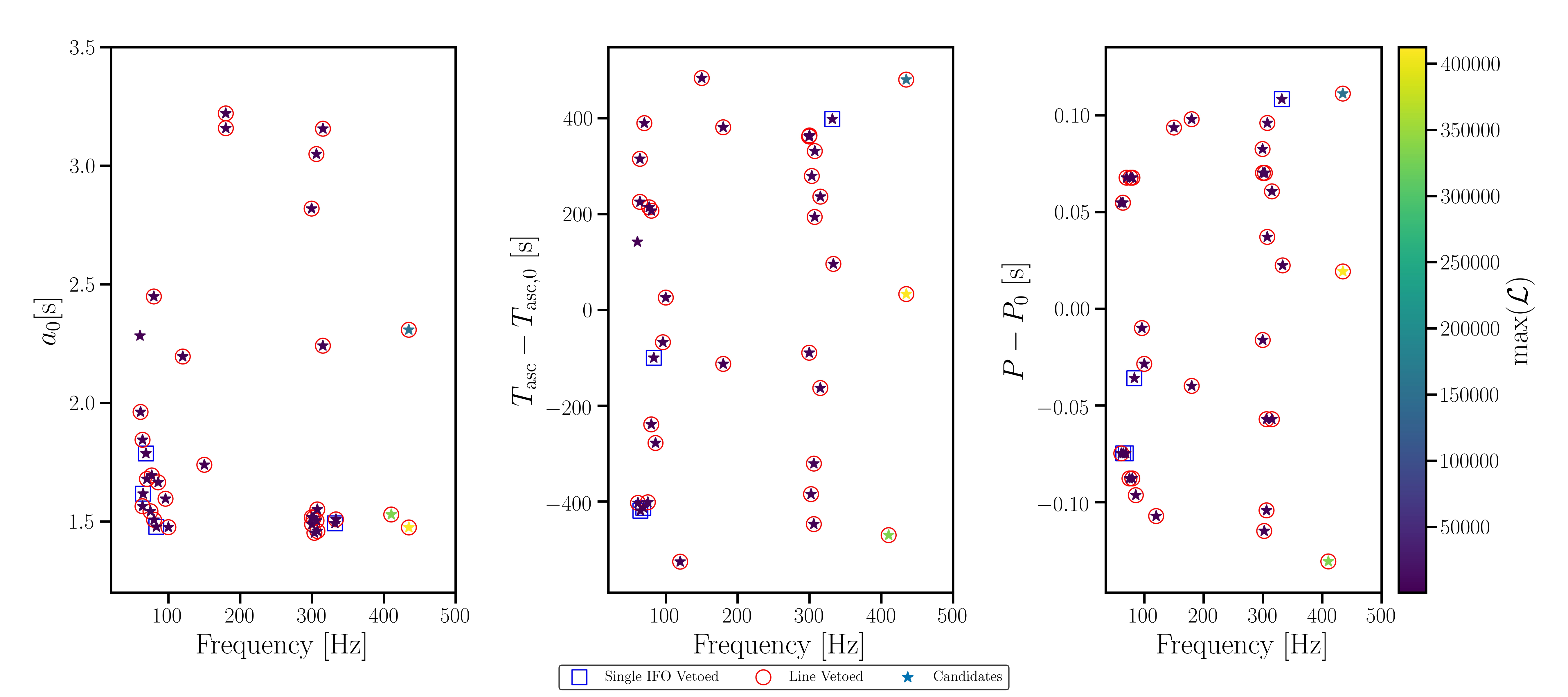}
    \caption{Candidates plotted as a function of their terminating frequency bin $q^{*}(t_{N_{T}})$ (horizontal axis, units in Hz) and the orbital parameters $a_{0}$ (vertical axis in left panel, units in s), offset from the central time of ascension $T_{\rm asc}-T_{\rm asc_{0}}$ (vertical axis in central panel; units in s) and offset from the central period $P-P_{0}$ (vertical axis in right panel; units in s). The color scale indicates the $\rm max(\mathcal{L})$ obtained for the candidate. Candidates marked with purple squares are eliminated  by the single IFO veto, while red circles mark the ones eliminated by the known lines veto. The candidate with no marking survives both the single IFO and known lines vetoes, but is eliminated by its absence when using noise-subtracted data (See Section \ref{subsec:Candidates}).}
    \label{fig:CandidatesperSub}
\end{figure*}

The results of the search are plotted in Figure \ref{fig:CandidatesperSub}. On the horizontal axis we show the terminating frequency bin of the optimal path, i.e $q^{*}(t_{N_{T}})$ per sub-band, that satisfies $\rm max(\mathcal{L}) > \mathcal{L}_{\text{th}}$, defining preliminary signal candidates. On the vertical axes we graph the orbital parameters $a_{0}$ (left panel), $T_{\text{asc}}$ (middle panel), and $P$ (right panel). 

Figure \ref{fig:CandidatesperSub} contains 35 candidates with $\rm max(\mathcal{L}) > \mathcal{L}_{\text{th}}$. To eliminate  false alarms we use the hierarchy of vetoes employed in Ref.~\cite{AbbottAbbott2019a}.  The vetoes discard candidates that (i) lie near an instrumental noise line (known line veto) or (ii) appear in one interferometer (IFO) but not the other (single IFO veto). In previous searches other vetoes, e.g. candidates that appear in half of the observation time ($T_{\rm obs}/2$ veto), have been used in addition to (i) and (ii) \cite{AbbottAbbott2017c,AbbottAbbott2019a,MiddletonClearwater2020,TheLIGOScientificCollaborationtheVirgoCollaboration2021a,TheLIGOScientificCollaborationtheVirgoCollaboration2021b}. Additional vetoes are unnecessary in this paper; all but one of the candidates are vetoed by (i) and (ii), and the survivor is eliminated by other means, as discussed below. The outcomes of the vetoes in the 35 sub-bands containing outliers are recorded in Table \ref{tab:Candidates}. 

The sole candidate that passes vetoes (i) and (ii), contained in the sub-band starting at $60.05$ Hz, is suspiciously close to the known $60$ Hz noise line due to the United States of America power grid \cite{AbbottAbbott2020}. Additionally, several other candidates appear near harmonics of $60.05$ Hz, for instance $119.52$ Hz and $179.60$ Hz. When we search the sub-band using the C01 calibrated self-gated 60 Hz subtracted dataset, which uses the algorithm described in Ref.~\cite{VajenteHuang2020} to subtract the $60$ Hz noise line, the candidate disappears.

\begin{table*}[tbp]
\caption{\label{tab:Candidates} Candidates for the O3 search. The first column corresponds to the minimum frequency in the sub-band that contains the candidate. The second column is the log likelihood of the candidate. The other columns record the outcome of the two vetoes used in Section \ref{veto:klv} and \ref{veto:single_IFO}. A candidate that passes or fails a veto is marked with a \checkmark or \textbf{X} respectively. H and L are the Hanford-only and Livingston-only $\rm max(\mathcal{L})$ values. The remaining candidate, contained in the $60.05$ Hz sub-band, is eliminated when using the ``C01 calibrated self-gated 60 Hz subtracted" dataset.}
\begin{ruledtabular}
\begin{tabular}{cccc}
Sub-band $(\text{Hz})$ & $\rm max(\mathcal{L})/10^{3}$ & Known lines veto & Single IFO veto \\ \hline 
60.05 &	0.45 &	\checkmark &	\checkmark; H:398.58, L:427.97    \\ 
60.66 &	0.40 &	\textbf{X} &	\textbf{-}  \\
63.09 &	1.73 &	\textbf{X} &	 \textbf{-}  \\ 	
63.70 &	1.39 &	\textbf{X} &	\textbf{-}  	\\ 
64.30 &	0.39 &	\checkmark &	\textbf{X}   \\ 
67.94 &	0.37 &	\checkmark &	\textbf{X}   \\ 
69.76 &	0.60 &	\textbf{X} &	\textbf{-}  \\ 
74.62 &	4.33 &	\textbf{X} &	\textbf{-}   \\ 
76.44 &	0.35 &	\textbf{X} &	\textbf{-}   \\ 
78.87 &	0.35 &	\textbf{X} &	\textbf{-}   \\ 
79.47 &	0.46 &	\textbf{X} &	\textbf{-}  \\ 
82.51 &	0.53 &	\checkmark & \textbf{X}  	\\ 
84.93 &	0.41 &	\textbf{X} &	\textbf{-}   \\ 
95.25 & 0.56 &	\textbf{X} &	\textbf{-}  \\ 
99.50 &	1.35 &	\textbf{X} &	\textbf{-} \\
119.52 & 0.64&	\textbf{X} &	\textbf{-}   \\ 
149.26 & 0.47&	\textbf{X} &	\textbf{-}  \\
178.99 & 0.43&	\textbf{X} &	\textbf{-}  \\ 
179.60 & 0.59&	\textbf{X} &	\textbf{-}  \\ 
298.53 & 1.18&	\textbf{X} &	\textbf{-}  \\
299.14 & 2.81&	\textbf{X} &	\textbf{-}  \\
299.75 & 0.38&	\textbf{X} &	\textbf{-}  \\
301.57 & 3.75&	\textbf{X} &	\textbf{-}  \\
302.78 & 4.26&	\textbf{X} &	\textbf{-}  \\
305.21 & 2.34&	\textbf{X} & \textbf{-}  \\
305.81 & 4.79&	\textbf{X} &	\textbf{-}  \\
306.42 & 0.59&	\textbf{X} &	\textbf{-} \\
307.03 & 5.21&	\textbf{X} &	\textbf{-}  \\
314.31 & 3.00&	\textbf{X} & \textbf{-} \\
314.92 & 1.50&	\textbf{X} & \textbf{-}  \\
331.30 & 6.40&	\checkmark & \textbf{X}  \\
332.51 & 0.45&	\textbf{X} & \textbf{-}  \\
409.58 & 332.67&	\textbf{X} &	\textbf{-}  \\
433.85 & 155.83&	\textbf{X}  & \textbf{-} \\
434.46 & 412.41&\textbf{X}  & \textbf{-} \\ \hline 
Total: 35 &  &	 &	  \\
\end{tabular}
\end{ruledtabular}
\end{table*}

\subsection{Vetoes}
\label{subsec_IV:Vetoes}

\subsubsection{Known lines}
\label{veto:klv}

Narrowband noise features in the IFO are caused by a plethora of reasons, such as the suspension system or the electricity grid \cite{AbbottAbbott2017d, CovasEffler2018}. As noise lines artificially increase the output of the $\mathcal{F}$-statistic, sub-bands that contain them tend to be flagged as candidates. In response, we veto any candidate whose optimal frequency path $f(t_{i})$ satisfies
\begin{equation}
    \vert f(t_{i}) - f_{\text{line}} \vert < \frac{2\pi a_{0} f_{\rm line}}{P},
    \label{eq:klv}
\end{equation}

\noindent for any epoch $t_{i}$ in the search. Here $f_{\text{line}}$ is the frequency of the noise line. We refer to the vetted known lines list in Ref.~\cite{O3lines}. This test vetoes 30 out of the 35 candidates. We note that the number of remaining candidates, after eliminating those caused by noise lines, is consistent with our original number of expected candidates.

\subsubsection{Single IFO}
\label{veto:single_IFO}

A plausible astrophysical signal that has escaped detection in prior searches would likely be weak enough to need data from both IFOs to be detectable, or strong  enough to be seen in both, given their comparable sensitivities in most frequency bands. In contrast, instrumental artifacts are unlikely to appear simultaneously in both IFOs. Let $\rm max(\mathcal{L})_{a}$ and $\rm max(\mathcal{L})_{b} > \rm max(\mathcal{L})_{a}$ denote the log likelihoods in each IFO, and let $\rm max(\mathcal{L})_{\cup}$ denote the log likelihood of the original candidate. There are four possible outcomes:

\begin{enumerate}
    \item If one finds $\rm max(\mathcal{L})_{a} < \mathcal{L}_{\rm th}$ and $\rm max(\mathcal{L})_{b} > max(\mathcal{L})_{\cup}$, and the optimal path, $f_{b}(t_{i})$, associated with $\rm max(\mathcal{L})_{b}$, satisfies
    \begin{equation}
        \vert f_{\cup}(t_{i})-f_{b}(t_{i}) \vert < \frac{2\pi a_{0} f_{\cup}}{P}
        \label{eq:single_IFO}
    \end{equation}
    
    for any epoch $t_{i}$ in the search, where $f_{\cup}(t_{i})$ is the optimal path associated with $\rm max(\mathcal{L})_{\cup}$ then the candidate is consistent with an instrumental artifact in detector $b$. It is vetoed.
    \item If one finds $\rm max(\mathcal{L})_{a} < \mathcal{L}_{\rm th}$ and $\rm max(\mathcal{L})_{b} > max(\mathcal{L})_{\cup}$ but the candidate does not satisfy Eq.~\eqref{eq:single_IFO}, then it is saved for further post-processing. Such a candidate could be a faint astrophysical signal which needs both IFOs to be detected. 
    
    \item If one finds $\rm max(\mathcal{L})_{a} > \mathcal{L}_{\rm th}$ and $\rm max(\mathcal{L})_{b} > \mathcal{L}_{\rm th}$, then the candidate could be a strong astrophysical signal. It could also imply a common noise source in both detectors. The candidate is flagged for post-processing.
    
    \item If one finds $\rm max(\mathcal{L})_{a} < \mathcal{L}_{\rm th}$ and $\rm max(\mathcal{L})_{b} < max(\mathcal{L})_{\cup}$, the candidate could be a weak astrophysical signal that needs both detectors to appear above threshold. This candidate is also saved for post-processing.
\end{enumerate}

This test vetoes four out of the remaining five candidates, with the last candidate eliminated by its non-appearance in noise-subtracted data, as discussed in Section \ref{subsec:Candidates}.

\section{Frequentist upper limits}
\label{sec_V: Upper limits}

\subsection{Procedure}

Sub-bands without candidates are used to place upper limits on the gravitational wave strain detectable at a $95\%$ confidence level, $h_{0}^{95\%}$. We use the approach in the historical HMM Sco X-1 searches \cite{AbbottAbbott2017c, AbbottAbbott2019a} to set frequentist upper limits. This is done to facilitate comparison with previous upper limits. 

To capture the variation of the wave strain as a function of the inclination angle $\iota$, we define the effective strain \cite{MessengerBulten2015}
\begin{equation}
    h^{\rm eff}_{0} = h_{0}2^{-1/2}\{\left[(1+\rm cos^{2} \iota)/2 \right]^{2} + \rm cos^{2} \iota \}^{1/2}.
    \label{Eq:eff_strain}
\end{equation}

We note that Eq.~\eqref{Eq:eff_strain} allows us to re-scale $h_{0}^{95 \%}$ for the circularly polarized case, $\rm \vert cos\, \iota \vert=1$, to any other inclination angle $\iota$. For example if we assume the electromagnetically measured inclination of Sco X-1 orbit as $\iota=44^{\circ}$ then Eq.~\eqref{Eq:eff_strain} yields $h_{0}^{{\rm eff}, 95\%}=1.35h_{0}^{95\%}$.

To set frequentist upper limits in a sub-band with central frequency $\bar{f}$, we start by generating 100 copies of the O3 data for this sub-band. A Sco X-1-like signal, i.e using the astrophysical parameters in Table \ref{tab:ScoX1Params}, is injected into each copy of the sub-band. The parameters $\{\psi,a_{0},T_{\rm asc},P\}_{\rm inj}$ used to create the injected signal are drawn from uniform distributions within the range given by their respective $3\sigma$ error bars. We make sure the injected $T_{\rm asc, inj}$ and $P_{\rm inj}$ values lie inside the propagated priors shown in Figure \ref{fig:propagation-Ellipses-example}, second panel. The injected frequency $f_{\rm inj}$ is uniformly selected from the interval $\bar{f}\pm \delta f$ with $\delta f=0.05$Hz. The interval $\bar{f}\pm \delta f$ is chosen for simplicity. The initial value of $h_{0}$ is chosen such that there is at least one frequency path with $\rm max(\mathcal{L}) > \mathcal{L}_{\rm th}$. We progressively reduce $h_{0}$, holding $\{\psi,a_{0},T_{\rm asc},P\}_{\rm inj}$ constant, until the signal is no longer detectable. We record the last detectable amplitude as $h_{0\, \rm min \, i}$. The procedure is repeated for all copies of the sub-band, choosing a new set of injection parameters $\{\psi,a_{0},T_{\rm asc},P\}_{\rm inj}$ per copy. Finally $\{h_{0\,\rm min\,1},...,h_{0\,\rm min\,100}\}$ are sorted in ascending order and the $95$th becomes $h_{0}^{95 \%}$. The injections have $\rm \vert cos \, \iota \vert=1$, so we use Eq.~\eqref{Eq:eff_strain} to convert $h_{0}^{95 \%}$ to other polarizations.

\subsection{Upper limits}

\begin{figure*}[ht]
    \centering
    \includegraphics[scale=0.5]{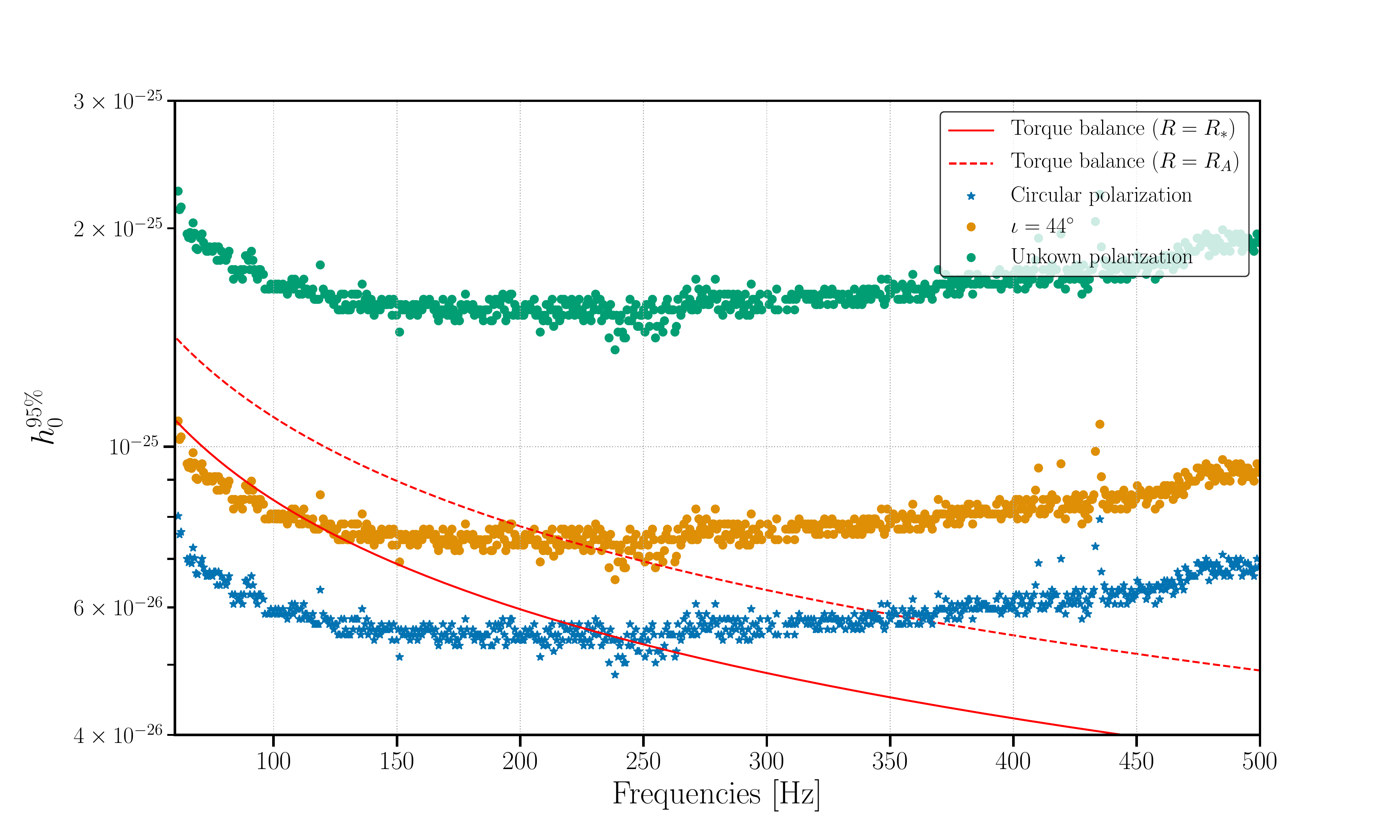}
    \caption{Frequentist effective wave strain upper limits at $95\%$ confidence as a function of sub-band frequency, for three scenarios: circular polarization with $\iota=0$ (blue stars), $\iota \approx 44^{\circ}$ based on the electromagnetic measurements (see Ref.~\cite{WangSteeghs2018}; orange dots), and a flat prior on $\rm cos\,\iota$ (green dots). Indirect torque-balance upper limits (see Section \ref{subsc_IV:Torque-balance upper limit}) for two torque lever arms are also shown: the stellar radius (red solid line) and the Alfv\'en radius (dashed red line).}
    \label{fig:UL_subbands}
\end{figure*}

\begin{figure*}[ht]
    \centering
    \includegraphics[scale=0.5]{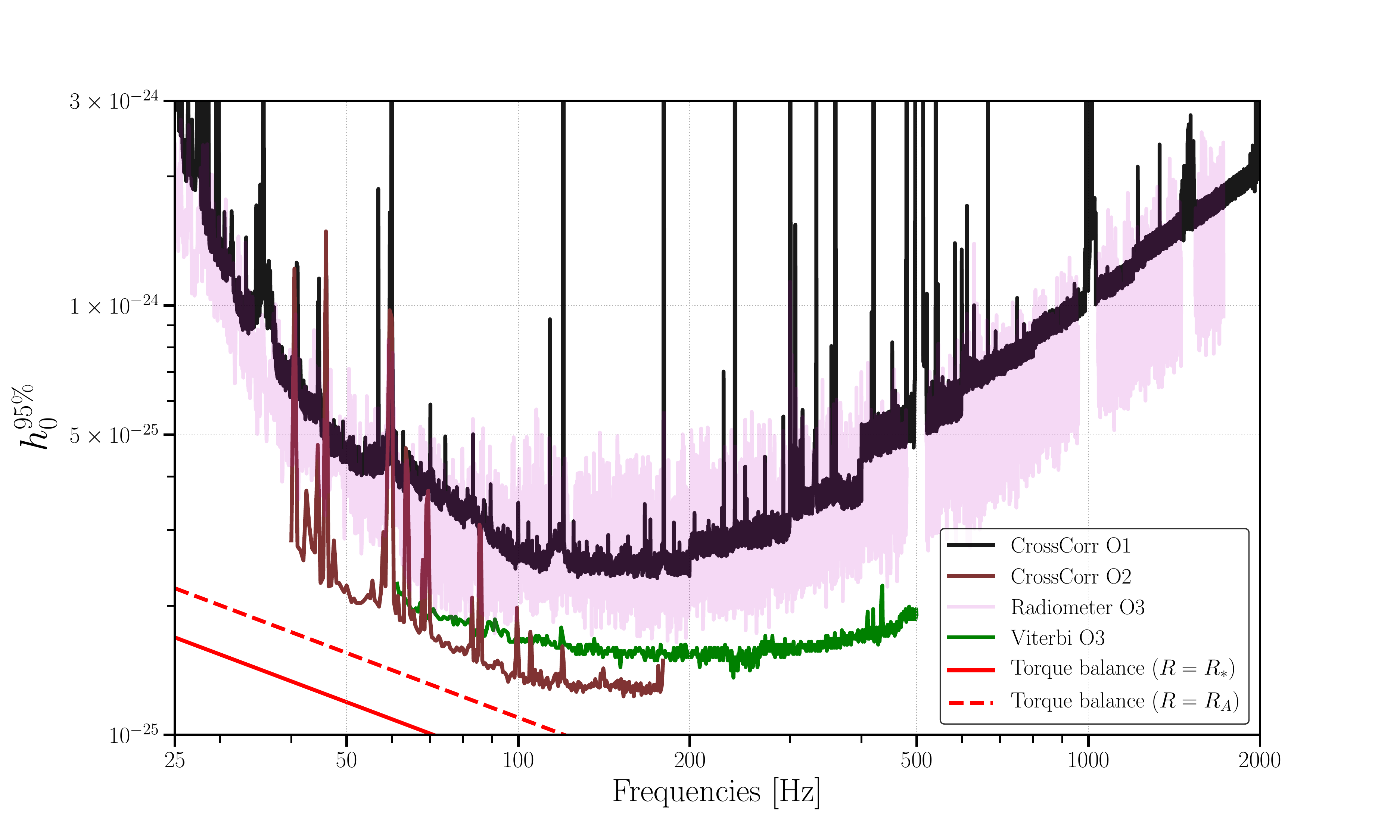}
    \caption{GW strain upper limits at $95\%$ confidence as a function of sub-band frequency, assuming  a uniform distribution in $\rm cos\,\iota$. The curves are for the CrossCorr O1 search (see Ref.~\cite{AbbottAbbott2017b}; black line), the CrossCorr O2 search (see Ref.~\cite{ZhangPapa2021}; brown line), the Radiometer O3 search (see Ref.~\cite{AbbottAbbott2021b}; light pink line), and the HMM search in this paper (green line). The indirect torque-balance upper limits (see Section \ref{subsc_IV:Torque-balance upper limit}), using the stellar radius (red solid line) and the Alfv\'en radius (dashed red line), are also plotted.}
    \label{fig:UL_comparison}
\end{figure*}

The limits on $h_{0}^{95 \%}$ are plotted in Figure \ref{fig:UL_subbands}. We present three cases, as in Ref.~\cite{AbbottAbbott2019a}: circular polarization $\iota=0$ (blue dots in Figure \ref{fig:UL_subbands}), $\iota=44^{\circ}$ (following the electromagnetic measurements in Ref.~\cite{WangSteeghs2018}; denoted by green dots in Figure \ref{fig:UL_subbands}), and unknown polarization (orange dots in Figure \ref{fig:UL_subbands}). In the latter context, unknown polarization means we marginalize over all possible polarizations assuming a uniform distribution in $\rm cos\,\iota$ from $-1$ to $1$. 

The upper limits from the O3 search are on average $\sim 3$ times lower than those from the O2 HMM search \cite{AbbottAbbott2019a}. 
For the sub-band starting at $256.06$Hz we obtain the lowest $h^{95\%}_{0}$, given by $4.56\times10^{-26}$, $6.16\times10^{-26}$, and $9.41\times10^{-26}$ for circular, electromagnetically restricted ($\iota=44^{\circ}$) and unknown polarizations, respectively. Compared to the most sensitive sub-bands in previous HMM Sco X-1 searches the lowest $h^{95\%}_{0}$ is a factor of $\sim 13$ lower than in O1 data \cite{AbbottAbbott2017c} and $\sim 3$ lower than in O2 data \cite{AbbottAbbott2019a}.
\vspace{-1pt}
\indent Figure \ref{fig:UL_comparison}, which is included for illustrative purposes, shows the GW upper limits marginalized over possible polarizations, for four different searches: CrossCorr O1 \cite{AbbottAbbott2017c} (black line), CrossCorr O2 \cite{ZhangPapa2021} (brown line), Radiometer O3 \cite{AbbottAbbott2021b} (light pink line),  and our results (green line). Each search is conditional on a different signal model, so the upper limits are not directly comparable. 

\subsection{Torque-balance upper limit}
\label{subsc_IV:Torque-balance upper limit}

Torque-balance assumes the spin-down torque due to gravitational wave emission balances the accretion spin-up torque. From this assumption theoretical upper limits on gravitational wave strain can be estimated from X-ray observations. Following Eq.~($4$) in Ref.~\cite{ZhangPapa2021}, the amplitude of the GW signal produced in torque equilibrium, $h^{\rm eq}_{0}$, is 
\begin{align}
h^{{\rm eq}}_{0}&=~3.4\times10^{-26}\left(\frac{R}{10 \rm km}\right)^{1/2}\left(\frac{1.4M_{\odot}}{M}\right)^{1/4}\left(\frac{r_{m}}{10 \rm km}\right)^{1/4} \nonumber\\
&\times\left(\frac{F_{X}}{3.9\times10^{-7} {\rm erg\,cm^{-2}s^{-1}}}\right)^{1/2} \left(\frac{600 \rm Hz}{f_{\rm GW}}\right)^{1/2}.
\label{Eq:ho_tb}
\end{align}
 
In Eq.~\eqref{Eq:ho_tb}, $F_{X}$ is the X-ray flux, $M$ is the fiducial neutron star mass, $r_{m}$ is the lever arm, $R$ is the neutron star radius and $f_{\rm GW}$ is the GW frequency. Eq.~\eqref{Eq:ho_tb} assumes the maximum accretion luminosity is completely radiated as X-rays, i.e. $X=1$ in Eq.~(4) of Ref.~\cite{ZhangPapa2021}. To calculate Eq.~\eqref{Eq:ho_tb}, we use $r_{m}=R=10$ km, plotted as a solid red line in Figure \ref{fig:UL_subbands}, or $r_{m}=R_{A}$, where $R_{A}$ is the Alfv\'en radius, which corresponds approximately to the inner edge of the accretion disk \cite{BildstenChakrabarty1997,AbbottAbbott2017c}. This is given by
\begin{align}
    R_{A} =& ~35\left(\frac{B_{\star}}{10^{9} G}\right)^{1/4}\left(\frac{R}{10 \rm km}\right)^{12/7} \nonumber\\
    &\times\left(\frac{1.4M_{\odot}}{M}\right)^{1/7}\left(\frac{10^{-8} M_{\odot} \rm yr^{-1}}{\dot{M}}\right)^{2/7} \rm km, 
    \label{eq:AlvenRad}
\end{align}

\noindent where $B_{\star}$ is the polar magnetic field strength at the stellar surface, $G$ is Newton's gravitational constant, and $\dot{M}$ is the accretion rate set to the Eddington limit $2\times10^{-8} M_{\odot} \rm yr^{-1}$, for a fiducial neutron star with mass $M=1.4M_{\odot}$ and radius $R=10$ km \cite{RitterKolb2003, SammutThesis}. This limit is plotted in Figure \ref{fig:UL_subbands} as the dashed red line.

The electromagnetic inclination $\iota=44^{\circ}$ produces upper limits that dip under the theoretical torque-balance limits (red lines in Figure \ref{fig:UL_subbands}) for the first time in the HMM Sco X-1 search history. The CrossCorr search pipeline achieved this milestone in the O2 search; see Figure 1 in Ref.~\cite{ZhangPapa2021}. Again, the reader is reminded that different pipelines assume different signal models, and upper limits conditional on different signal models cannot be compared directly.

\section{Conclusions}
\label{sec_VI:Conclusions}

In this paper we search the LIGO O3 data for continuous GWs from the LMXB Sco X-1, using a hidden Markov model, combined with the maximum-likelihood $\mathcal{J}$-statistic and a binary template grid that includes the projected semi major axis $a_{0}$, time of ascension $T_{\rm asc}$ and orbital period $P$. The binary orbital elements are constrained via electromagnetic observations, but the spin frequency is unknown. Monte-Carlo simulations are used to establish a detection threshold, $\mathcal{L}_{\rm th}$, with a false alarm probability $\alpha_{N_{\rm tot}}=0.01$, per sub-band. The search is conducted in the range $60$Hz to $500$Hz, partitioned into sub-bands of width $\Delta f_{\rm sub}= 0.606$Hz. The sub-bands with an optimal path satisfying $\rm max(\mathcal{L})>\mathcal{L}_{\rm th}$ are passed through a hierarchy of vetoes. One candidate survives the vetoes, but this candidate is eliminated when using the ``C01 calibrated self-gated 60 Hz subtracted" dataset.

The most sensitive sub-band, starting at $256.06$Hz, yields $h^{95\%}_{0}=4.56\times10^{-26}, 6.16\times10^{-26},$ and $9.41\times10^{-26}$ for circular, electromagnetically restricted ($\iota=44^{\circ}$) and unknown polarizations, respectively.

The above results improve on the two previous HMM Sco X-1 searches \cite{AbbottAbbott2017c,AbbottAbbott2019a} by using data from O3 and including the orbital period $P$ in the searched template grid. For comparison, the most sensitive sub-band in the O2 HMM search, $194.6$ Hz, obtained $h_{0}^{95\%}=1.42\times10^{-25}$ for $\iota=0$ \cite{AbbottAbbott2019a}, while for the same sub-band and polarization the present search obtains $h_{0}^{95\%}=5.40\times10^{-26}$. On average our upper limits are a factor of $\sim 3$ below the O2 HMM results. The present search sets the lowest upper-limits for the HMM searches, beating for first time the torque-balance limit for the electromagnetically restricted $\iota=44^{\circ}$ case. 

Other LMXBs are not as bright in X-rays as Sco X-1, but they are important targets too. Some LMXBs emit X-ray pulsations, so that $f_{\star}$ is measured to high precision electromagnetically, an important advantage. However the gravitational wave frequency emitted by such objects may be displaced from $f_{\star}$ and wander randomly with respect to it. An HMM-based search is well placed to track such wandering. Searches for LMXBs with electromagnetically-constrained rotation frequencies have been performed in O2 \cite{MiddletonClearwater2020} and O3 \cite{TheLIGOScientificCollaborationtheVirgoCollaboration2021a} data. Ref.~\cite{TheLIGOScientificCollaborationtheVirgoCollaboration2021a} reported strain upper limits in the range $5.1\times10^{-26} \leq h_{0}^{95\%} \leq 1.1\times10^{-25} $ for its $20$ candidates. Such searches offer considerable promise in future observing runs. 

\section{Acknowledgments}

This material is based upon work supported by NSF’s LIGO Laboratory which is a major facility
fully funded by the National Science Foundation.
The authors also gratefully acknowledge the support of
the Science and Technology Facilities Council (STFC) of the
United Kingdom, the Max-Planck-Society (MPS), and the State of
Niedersachsen/Germany for support of the construction of Advanced LIGO 
and construction and operation of the GEO\,600 detector. 
Additional support for Advanced LIGO was provided by the Australian Research Council.
The authors gratefully acknowledge the Italian Istituto Nazionale di Fisica Nucleare (INFN),  
the French Centre National de la Recherche Scientifique (CNRS) and
the Netherlands Organization for Scientific Research (NWO), 
for the construction and operation of the Virgo detector
and the creation and support  of the EGO consortium. 
The authors also gratefully acknowledge research support from these agencies as well as by 
the Council of Scientific and Industrial Research of India, 
the Department of Science and Technology, India,
the Science \& Engineering Research Board (SERB), India,
the Ministry of Human Resource Development, India,
the Spanish Agencia Estatal de Investigaci\'on (AEI),
the Spanish Ministerio de Ciencia e Innovaci\'on and Ministerio de Universidades,
the Conselleria de Fons Europeus, Universitat i Cultura and the Direcci\'o General de Pol\'{\i}tica Universitaria i Recerca del Govern de les Illes Balears,
the Conselleria d'Innovaci\'o, Universitats, Ci\`encia i Societat Digital de la Generalitat Valenciana and
the CERCA Programme Generalitat de Catalunya, Spain,
the National Science Centre of Poland and the European Union – European Regional Development Fund; Foundation for Polish Science (FNP),
the Swiss National Science Foundation (SNSF),
the Russian Foundation for Basic Research, 
the Russian Science Foundation,
the European Commission,
the European Social Funds (ESF),
the European Regional Development Funds (ERDF),
the Royal Society, 
the Scottish Funding Council, 
the Scottish Universities Physics Alliance, 
the Hungarian Scientific Research Fund (OTKA),
the French Lyon Institute of Origins (LIO),
the Belgian Fonds de la Recherche Scientifique (FRS-FNRS), 
Actions de Recherche Concertées (ARC) and
Fonds Wetenschappelijk Onderzoek – Vlaanderen (FWO), Belgium,
the Paris \^{I}le-de-France Region, 
the National Research, Development and Innovation Office Hungary (NKFIH), 
the National Research Foundation of Korea,
the Natural Science and Engineering Research Council Canada,
Canadian Foundation for Innovation (CFI),
the Brazilian Ministry of Science, Technology, and Innovations,
the International Center for Theoretical Physics South American Institute for Fundamental Research (ICTP-SAIFR), 
the Research Grants Council of Hong Kong,
the National Natural Science Foundation of China (NSFC),
the Leverhulme Trust, 
the Research Corporation, 
the Ministry of Science and Technology (MOST), Taiwan,
the United States Department of Energy,
and
the Kavli Foundation.
The authors gratefully acknowledge the support of the NSF, STFC, INFN and CNRS for provision of computational resources.
This work was supported by MEXT, JSPS Leading-edge Research Infrastructure Program, JSPS Grant-in-Aid for Specially Promoted Research 26000005, JSPS Grant-in-Aid for Scientific Research on Innovative Areas 2905: JP17H06358, JP17H06361 and JP17H06364, JSPS Core-to-Core Program A. Advanced Research Networks, JSPS Grant-in-Aid for Scientific Research (S) 17H06133 and 20H05639 , JSPS Grant-in-Aid for Transformative Research Areas (A) 20A203: JP20H05854, the joint research program of the Institute for Cosmic Ray Research, University of Tokyo, National Research Foundation (NRF), Computing Infrastructure Project of KISTI-GSDC, Korea Astronomy and Space Science Institute (KASI), and Ministry of Science and ICT (MSIT) in Korea, Academia Sinica (AS), AS Grid Center (ASGC) and the Ministry of Science and Technology (MoST) in Taiwan under grants including AS-CDA-105-M06, Advanced Technology Center (ATC) of NAOJ, and Mechanical Engineering Center of KEK

\appendix

\bibliography{only_ads_references_II,only_non_ads_references}
\end{document}